\documentclass{article}

\usepackage{arxiv}
\usepackage{array} 
 \newcolumntype{L}{>{\centering\let\newline\\\arraybackslash\hspace{0pt}}m{2cm}}
\usepackage{threeparttable} % Add this line in the preamble
\usepackage{multirow}
\usepackage{epsfig}
\usepackage{natbib}
\usepackage{amsmath}
\usepackage[utf8]{inputenc} % allow utf-8 input
\usepackage[T1]{fontenc}    % use 8-bit T1 fonts
\usepackage{url}            % simple URL typesetting
\usepackage{booktabs}       % professional-quality tables
\usepackage{amsfonts}       % blackboard math symbols
\usepackage{nicefrac}       % compact symbols for 1/2, etc.
\usepackage{microtype}      % microtypography
\usepackage{lipsum}
\usepackage{graphicx}
\graphicspath{ {./images/} }
\def\bSig\mathbf{\Sigma}

\usepackage[table]{xcolor}
\usepackage{amsmath}
\usepackage{amsfonts}
\usepackage{array}
\usepackage{placeins}
\usepackage{moreverb}
\usepackage{bm}
\usepackage{booktabs}
\usepackage[colorlinks=true,linkcolor=red,citecolor=blue]{hyperref}
\newtheorem{proposition}{Proposition}

\title{A Bayesian adaptive enrichment design using aggregate historical data to inform individualized treatment recommendations}

\author{
 Lara Maleyeff \\
 Montreal Heart Institute \\ Department of Medicine, Université de Montréal \\ Montr\'eal, QC, CA \\
  \texttt{lara.maleyeff@mcgill.ca} \\
   \And
 Shirin Golchi \\
 Department of Epidemiology, Biostatistics, and Occupational Health \\ McGill University, Montr\'eal, QC, CA \\
  \And
 Erica E. M. Moodie \\
Department of Epidemiology, Biostatistics, and Occupational Health \\ McGill University, Montr\'eal, QC, CA
}

\begin{document}
\maketitle
\begin{abstract}
Adaptive enrichment trials aim to identify and recruit participants most likely to benefit from treatment based on evolving biomarker evidence, with the goal of informing individualized treatment recommendations. Bayesian methods are well suited to these designs because they allow external information to be incorporated in a principled manner. In practice, prior studies often provide only summary-level information, with subgroup-specific estimates unavailable due to design or privacy constraints. Existing dynamic borrowing approaches therefore rely on aggregate measures, such as the average treatment effect, and implicitly assume that historical information maps directly onto model parameters. In adaptive enrichment settings aimed at identifying individualized treatment effects, however, subgroup-specific treatment parameters are not identifiable when only marginal historical effects are available. To address this gap, we propose a Bayesian adaptive enrichment design that borrows information from external studies using a normalized power prior anchored on one or more summary measures, such as the average treatment effect. { To our knowledge, no existing method addresses this gap.} Interim analyses use posterior probabilities to guide early stopping for efficacy or futility, or to continue recruitment within promising biomarker-defined subgroups. Simulation studies evaluate operating characteristics across historical bias, sample size, and prior informativeness. Together with a motivating future trial in obstructive sleep apnea, the results show efficiency gains versus non-borrowing designs, including improved power, earlier stopping, and reduced expected sample size.
\end{abstract}

% keywords can be removed


\section{Introduction}\label{sec:intro}

Precision medicine promises improved patient outcomes by tailoring experimental treatment and prevention strategies to an individual's unique biomarker profile \citep{thall2024bayesian}. Randomised controlled trials (RCTs), the gold standard for regulatory decision-making, are typically powered to estimate an average treatment effect (ATE) in a prespecified population. Consequently, most conventional RCTs are underpowered to detect clinically meaningful effect heterogeneity.  For example, in obstructive sleep apnea (OSA), recent work by Azarbarzin and colleagues has shown that two oximetry-derived biomarkers-hypoxic burden (HB) and the heart rate response to respiratory events-predict cardiovascular risk and modify response to positive airway pressure (PAP) therapy \citep{azarbarzin2019hypoxic,azarbarzin2021sleep}. Patients with high biomarker values appear to benefit from PAP treatment, whereas those with low values experience little or no benefit, illustrating substantial treatment-subgroup interactions that conventional trial designs may fail to detect.  Such challenges motivate the structured borrowing of external information to strengthen subgroup analyses. As only summary-level information is often available from prior studies, and subgroup-specific estimates are frequently lacking or not reported due to design or privacy restrictions, we focus on borrowing from summary measures, such as the ATE, rather than subgroup-specific effects. To this end, we propose extending the normalized power prior (NPP) in order to borrow adaptively from historical summary measures, thereby improving power for subgroup analyses while guarding against prior-data conflict and overcoming data availability limitations.

Adaptive enrichment designs are an efficient class of biomarker-driven trials that begin with broad eligibility and, based on accumulating evidence, prospectively restrict or expand enrollment to patient subgroups that appear most likely to benefit. These designs can improve power, ethical balance, and resource use relative to fixed designs \citep{simon2013adaptive}. Because decisions depend on evolving evidence, Bayesian methods are particularly suited: posterior probabilities yield directly interpretable interim decision criteria, predictive probabilities quantify the chance of ultimate trial success, and Bayesian hierarchical models allow information to be borrowed across related subgroups or external sources in a principled way \citep{giovagnoli2021bayesian}. Dynamic borrowing methods ensure that borrowing is attenuated in the presence of prior-data conflict, maintaining validity even under partial discordance.  Methods for identifying subgroups either assume pre-specified subgroup definitions \citep{simon2013adaptive, freidlin2013phase} or adaptively discover subgroups based on accumulating data \citep{park2022bayesian,maleyeff2024adaptive}. Interim decisions are prespecified and typically include early stopping for efficacy or futility and rules for enrichment; these interim analyses and decision procedures can be implemented with frequentist group-sequential monitoring using spending functions or pre-specified posterior cutoffs to maintain error control \citep{pocock1977group,stallard2020comparison}. 
%A related design is the biomarker-driven basket trial, which evaluates a single intervention across multiple diseases or molecularly defined subtypes in parallel and often incorporates information sharing across baskets. Unlike enrichment within one indication, basket trials determine where a therapy is active across indications rather than which subset within an indication should continue to be enrolled \citep{redig2015basket}. Bayesian hierarchical models are often employed to share information adaptively across baskets while down-weighting subgroups that are not commensurate, thereby improving efficiency without compromising validity \citep{psioda2021bayesian}.

{Although strict control of the frequentist Type I error rate is theoretically unattainable when borrowing from external data \citep{kopp2020power}, efforts must be made to minimize inflation. The U.S. Food and Drug Administration (FDA) guidance on adaptive designs specifies that Bayesian designs incorporating informative priors must evaluate ``the chances of erroneous conclusions, including the chances of false positive conclusions, under various scenarios of prior-data conflict" through pre-specified simulation studies \citep{fda2025adaptive}.} Careful design can further minimize inflation through conservative modeling choices and robust prior formulations.  Bayesian dynamic borrowing methods adaptively down-weight discordant information, ensuring that current trial evidence predominates in the presence of prior-data conflict. Among these, we focus on the power-prior family for its well-documented advantages: historical evidence enters as a likelihood raised to a weight, which makes borrowing transparent and tunable, the structure preserves standard likelihood properties (including straightforward propriety checks and familiar asymptotics), and prior evidence is discounted when current and historical information diverge \citep{ibrahim2015power}. The NPP further allows the weight to be learned while maintaining propriety via a normalizing constant \citep{duan2005modified,duan2006evaluating,carvalho2021normalized}. Using informative discounted priors offers practical advantages over other ways of incorporating external information--such as direct augmentation with external controls, synthetic controls, or unconditional hierarchical pooling--because discounting provides built-in protection against lack of exchangeability, keeps the primary likelihood for the randomized data explicit, and supports modular use of either patient-level or summary-level historical inputs. Closely related approaches include commensurate priors, which adapt borrowing through a similarity variance, supervised priors, which calibrate informativeness using compatibility metrics, and robust meta-analytic-predictive (MAP) priors, which mix a weakly informative component with a meta-analytic prior to balance efficiency and robustness \citep{hobbs2012commensurate,pan2017calibrated,schmidli2014robust}. { These methods require historical information on the same parameters targeted by the current model.}

Recent work has begun to address subgroup analyses using external information in contexts where patient-level historical data is available. \cite{schwartz2023harmonized} study RCTs augmented with patient-level external controls and introduce harmonized estimators that first estimate subgroup effects using RCT and external-control data and then adjust them so their prevalence-weighted average matches the RCT-only overall effect, ensuring coherence with the primary analysis. \cite{Ventz2025BOEED} develop a Bayesian optimal enrichment framework that jointly models trial data and concurrent external control data and computes stage-wise enrichment and stopping rules via dynamic programming-again assuming access to individual-level external controls. 

In this article, we develop a Bayesian adaptive enrichment design that borrows external evidence through a NPP anchored on one or more summary measures to inform individualized treatment recommendations. The contribution of this work is two-fold. Firstly, we extend the NPP to settings where only summary-level historical information is available by mapping under-identified external summaries to model parameters via a general function, with closed-form results for linear links and a Taylor-expansion-based approximation method for nonlinear links (Section~\ref{s:npp}). { To our knowledge, no existing dynamic borrowing 
method accommodates this common setting, as current approaches either require 
patient-level historical data or historical information on the same parameters 
targeted by the current model.} Secondly, we integrate this prior with a prospectively specified adaptive enrichment framework that identifies an effective subspace, applies posterior-probability rules for early efficacy or futility, and restricts subsequent enrollment to promising subgroups (Sections~\ref{s:trial_design}-\ref{s:interim}). We evaluate finite-sample validity and efficiency through simulation studies that compare the borrowing design with a no-borrowing reference across various realistic settings (Section~\ref{sec:sims}). We illustrate the proposed method using a future adaptive enrichment trial in OSA designed to detect treatment-sensitive subgroups using information from multiple historical studies (Section~\ref{sec:data_example}). Practical guidance, limitations, and avenues for extension are discussed in Section~\ref{sec:discussion}.

\section{Methods}
\label{s:methods}

\subsection{Probability Model}
\label{s:prob_model}

We assume outcomes $\{Y_i\}_{i=1}^n$ from the current trial arise from a regression model with a treatment-covariate interaction. Let $\mu_i = \mathbb{E}[Y_i \mid \mathrm{T}_i, X_i]$ denote the conditional mean (or other scale parameter of interest), linked to covariates through
\begin{equation}
    \eta_i = g(\mu_i) \;=\; \beta_0 + \beta_1 X_i + \beta_2 \,\mathrm{T}_i + \beta_3 \,\mathrm{T}_i X_i,
\end{equation}
where $g(\cdot)$ is an appropriate link function, $\mathrm{T}_i \in \{0,1\}$ is the treatment indicator, and $X_i$ is a baseline covariate. This formulation covers generalized linear models for continuous, binary, or count outcomes, as well as survival models such as the proportional hazards model, where $g(\mu_i)$ is the log-hazard. In a Bayesian framework the data are treated as fixed, so the model is not restricted to exponential family distributions; any likelihood paired with the linear predictor may be used. { We note that the historical study is not required to share the same model structure as the current trial; the proposed framework requires only that the historical study reports a summary measure that can be expressed as a known function of the current model parameters, as described in Section~\ref{s:npp}. The historical covariate distribution, captured through the biomarker prevalence, need not match that of the current trial and may be derived from the historical study's reported summary statistics.

We focus throughout on a single binary biomarker $X_i \in \{0,1\}$. While designs that consider multiple biomarkers simultaneously exist---including basket trials and multi-marker subgroup identification approaches---such designs require substantially larger sample sizes to achieve adequate power across a higher-dimensional covariate space. The single-biomarker enrichment setting is therefore common in practice when sample sizes are limited \citep{simon2013adaptive, freidlin2013phase}, and serves as a natural and practically motivated starting point for the present framework. 
}

\subsection{Normalized Power Prior Framework}
\label{s:npp}

To incorporate historical information, we adopt the NPP \citep{duan2005modified,duan2006evaluating,carvalho2021normalized}. The standard (unnormalized) power prior raises the historical likelihood to a fractional power $a$ and multiplies it by the baseline prior, but this can yield an improper posterior if $a$ is treated as unknown. The NPP resolves this issue by including a normalizing constant $C(a)$, ensuring a proper joint prior for $(\boldsymbol{\beta},a)$ and making it possible to place a hyperprior on $a$. 

Let $\boldsymbol{\beta}$ denote the full vector of model parameters, and let $\pi_0(\boldsymbol{\beta})$ represent the baseline prior, typically weakly informative or non-informative. Suppose external evidence $\mathcal{D}_0$ provides information on a function of a subset of parameters rather than on the parameters themselves. Denote this function as $\boldsymbol{\Delta} = h(\boldsymbol{\beta}_E),$ where $\boldsymbol{\beta}_E \subseteq \boldsymbol{\beta}$ and $h(\cdot)$ is a known mapping from the relevant subset of parameters to a vector of summaries, such as ATEs, log-odds ratios, or other linear and nonlinear combinations of regression coefficients. The dimension of $\boldsymbol{\Delta}$ may be one or greater, depending on the available summaries.

The NPP augments the baseline prior with a fractional contribution of the historical likelihood,
\begin{equation}
\label{eq:npp}
\pi(\boldsymbol{\beta},a \mid \mathcal{D}_0) \;\propto\; 
\frac{L_{\text{sum}}(h(\boldsymbol{\beta}_E))^{a}}{C(a)} \;\pi_0(\boldsymbol{\beta})\,\pi(a),
\end{equation}
where $L_{\text{sum}}(h(\boldsymbol{\beta}_E))$ is the likelihood derived from external summaries, $a\in[0,1]$ is a penalty parameter that governs the effective weight given to historical information, and $\pi(a)=\mathrm{Beta}(\eta,\nu)$ provides adaptive shrinkage through a hyperprior. The normalizing constant
\begin{equation}
\label{eq:c_true}
C(a)=\int L_{\text{sum}}(h(\boldsymbol{\beta}_E))^a \pi_0(\boldsymbol{\beta})\,d\boldsymbol{\beta}
\end{equation}
ensures that the NPP integrates to one, and its closed form (when available) facilitates efficient computation. If the external summary is approximately multivariate Normal, $\boldsymbol{\Delta} = h(\boldsymbol{\beta}_E) \sim \mathcal{N}(m_\Delta, \Sigma_\Delta)$, then the corresponding summary-likelihood is
\begin{equation} 
\label{eq:l_summ}
L_{\text{sum}}\big(h(\boldsymbol{\beta}_E)\big) \propto
\exp\!\left[-\tfrac{1}{2}
\big(h(\boldsymbol{\beta}_E) - m_\Delta\big)^\top
\Sigma_\Delta^{-1}
\big(h(\boldsymbol{\beta}_E) - m_\Delta\big)
\right].
\end{equation}
This form applies whether $h(\cdot)$ is linear or nonlinear.

In practice, the NPP can be implemented in \texttt{Stan} using Hamiltonian Monte Carlo (a gradient-based Markov chain Monte Carlo [MCMC] method), to sample from the posterior. This requires evaluation of the joint log-density and its gradients with respect to $(\boldsymbol{\beta},a)$. Because gradients propagate through $h(\boldsymbol{\beta}_E)$ and through the summary-likelihood term, the sampler automatically handles the induced borrowing structure without additional user-specified algorithms. When $h(\cdot)$ is linear or represented through a Taylor expansion, $\log C(a)$ has a closed form and can be passed to \texttt{Stan} as data; otherwise, a Monte Carlo approximation may be computed externally and supplied as a lookup table.

\subsubsection{Linear $h(\cdot)$ Case}

When both the baseline prior and the summary-likelihood are Gaussian, and $h(\boldsymbol{\beta}_E)$ is linear in the parameters, the NPP admits closed-form expressions. Assume
\begin{equation}
\pi_0(\boldsymbol{\beta}) = \mathcal{N}(m_0, \Sigma_0),
\quad
h(\boldsymbol{\beta}_E) = D \boldsymbol{\beta}_E,
\end{equation}
where $D$ is a known matrix defining the linear contrasts corresponding to the external summaries. Then
\begin{equation}
L_{\text{sum}}\big(h(\boldsymbol{\beta}_E)\big) \propto
\exp\!\left[-\tfrac{1}{2}
\big(D \boldsymbol{\beta}_E - m_\Delta\big)^\top
\Sigma_\Delta^{-1}
\big(D \boldsymbol{\beta}_E - m_\Delta\big)
\right],
\end{equation}
and the normalizing constant has a closed form:
{\small
\begin{equation}
C(a) =
\big| I + a \Sigma_0^{1/2} D^\top \Sigma_\Delta^{-1} D \Sigma_0^{1/2} \big|^{-1/2}
\exp\!\left[-\tfrac{1}{2}
a^2 (m_\Delta - D m_0)^\top
\Sigma_\Delta^{-1/2}
\big(I + a \Sigma_\Delta^{-1/2} D \Sigma_0 D^\top \Sigma_\Delta^{-1/2}\big)^{-1}
\Sigma_\Delta^{-1/2}
(m_\Delta - D m_0)
\right].
\end{equation}
}
This case includes, for example, linear regression models with identity link functions or any setting where the external summaries can be represented as linear combinations of regression coefficients.

\subsubsection{Nonlinear $h(\cdot)$ and Complex Treatment Effect Heterogeneity}
\label{s:nonlinear}

Nonlinear mappings $h(\boldsymbol{\beta}_E)$ arise frequently when external summaries are reported on scales such as marginal log-odds ratios (logistic models), rate ratios (Poisson models), or hazard ratios (survival models). In these settings the summary-likelihood is no longer linear in $\boldsymbol{\beta}_E$, and the NPP normalizing constant $C(a)$ has no closed-form expression. We propose a practical approximation to the mapping using a first-order Taylor expansion around a reference value $\boldsymbol{\beta}^\star_E$, which can be taken as the non-borrowing maximum likelihood estimator (MLE) from the current trial:
\begin{equation}
\label{eq:h}
h(\boldsymbol{\beta}_E)
\approx 
\tilde{h}(\boldsymbol{\beta}_E)=h(\boldsymbol{\beta}^\star_E) +
J(\boldsymbol{\beta}^\star_E)
(\boldsymbol{\beta}_E - \boldsymbol{\beta}^\star_E),
\end{equation}
where $J(\boldsymbol{\beta}^\star_E) = \partial h / \partial \boldsymbol{\beta}_E^\top \big|_{\boldsymbol{\beta}_E=\boldsymbol{\beta}_E^*}$ is the Jacobian evaluated at $\boldsymbol{\beta}_E^*$. Substituting this linearization into $L_{\mathrm{sum}}(h(\boldsymbol{\beta}_E))$ yields a Gaussian approximation and permits reuse of the closed-form expression for $C(a)$ from the linear case. Table~\ref{tab:mapping-table} provides explicit forms of $h(\boldsymbol{\beta}_E)$ and its Jacobian for several common link functions. { Web Appendix~A confirms the accuracy of the first-order Taylor expansion in the presence of non-collapsibility, and the simulation study in Section~\ref{sec:sims} 
further corroborates this numerically: the Linearized and Exact implementations 
produce nearly indistinguishable operating characteristics across all simulation scenarios examined. The following proposition formalizes the validity of this approximation, showing that the linearized and exact NPPs are asymptotically equivalent as the current trial sample size grows.

\begin{proposition}[Validity of the Linearized NPP]
\label{prop:taylor}
Let $\tilde{h}(\boldsymbol{\beta}_E)$ be the first-order Taylor expansion of $h$ defined in~\eqref{eq:h}, expanded around $\boldsymbol{\beta}_E^* = \hat{\boldsymbol{\beta}}_{E,n}$, MLE subvector from $\mathcal{D}_n$. Define the full posterior log-density error
\[
\mathcal{E}(\boldsymbol{\beta}_E, a) :=
\bigl[a\log L_{\mathrm{sum}}(h(\boldsymbol{\beta}_E)) - \log C(a)\bigr]
-
\bigl[a\log L_{\mathrm{sum}}(\tilde{h}(\boldsymbol{\beta}_E))
- \log \tilde{C}(a)\bigr],
\]
where $\tilde{C}(a) = \int L_{\mathrm{sum}}(\tilde{h}(\boldsymbol{\beta}_E))^a
\pi_0(\boldsymbol{\beta})\,d\boldsymbol{\beta}$ is the normalizing constant of the linearized NPP, and $C(a)$ and $L_{\mathrm{sum}}$ are as defined
in~\eqref{eq:c_true} and~\eqref{eq:l_summ} respectively. Let $\Pi$ denote the exact joint NPP posterior
\[
\pi(\boldsymbol{\beta}, a \mid \mathcal{D}_0, \mathcal{D}_n)
\;\propto\;
\frac{L_{\mathrm{sum}}\bigl(h(\boldsymbol{\beta}_E)\bigr)^a}{C(a)}
\,L_n(\boldsymbol{\beta})\,\pi_0(\boldsymbol{\beta})\,\pi(a),
\]
where $L_n(\boldsymbol{\beta})$ is the current-trial likelihood and let $P_0$ denote its true data-generating distribution. Assume the standard regularity conditions (B.1-B.5 in Web Appendix B) for the Bernstein--von Mises theorem hold for $L_n$. Then
\[
\mathbb{E}_{\Pi}\!\left[|\mathcal{E}(\boldsymbol{\beta}_E, a)|\right]
= O_{P_0}(n^{-1}).
\]
Proof provided in Web Appendix B.
\end{proposition}

This result provides theoretical justification for using the linearized NPP throughout the remainder of the paper. Two structural features of the mapping are worth noting. First, $h(\boldsymbol{\beta}_E)$ does not uniquely identify all elements of $\boldsymbol{\beta}_E$, but this is by design: the NPP uses $h(\boldsymbol{\beta}_E)$ to inform rather than fully determine the posterior, with the remaining information coming from the current trial. Second, in the single binary biomarker setting the model is saturated and $h(\boldsymbol{\beta}_E)$ is a deterministic function of the model parameters; the only potential source of misspecification is the historical biomarker prevalence $\mu_X^{\text{hist}}$, which enters explicitly into Table~\ref{tab:mapping-table} and must be obtained from the historical study. We recommend verifying consistency of the definition of $X$ across studies and conducting sensitivity analyses over plausible values of $\mu_X^{\text{hist}}$ when there is uncertainty. Web Appendix~C extends these mappings to the case of a general categorical 
biomarker with $K$ levels, providing closed-form expressions for the 
identity-identity and logit-logit cases that reduce to Table~\ref{tab:mapping-table} when $K=2$.}

When treatment effect heterogeneity is expressed as a smooth function of a continuous biomarker using basis expansions $f(x)=\sum_{k=1}^K b_k B_k(x)$, the mapping $h(\boldsymbol{\beta}_E)$ links the external summary (often an ATE-like functional of $f$) to the spline coefficients. If the historical estimand is linear in $f$, such as 
$\theta = \int f(x)\,dF_X(x)=\sum_{k} b_k \bar B_k$, 
then the summary-likelihood is approximately Gaussian and the closed-form expression for $C(a)$ from the linear-Gaussian case applies directly with $D=\bar{\mathbf{B}}^\top$. When the estimand is nonlinear--for example, a marginal risk or odds-ratio contrast induced by $f$--the mapping becomes curved and no closed-form $C(a)$ exists. In this setting, the Taylor-based linearization described above extends naturally: one expands the nonlinear functional $h(\boldsymbol{\beta}_E)$ around $\boldsymbol{\beta}^\star_E$, substitutes the linear approximation into the summary-likelihood, and thereby recovers a Gaussian working model with an analytic expression for $C(a)$. This provides a principled and scalable way to incorporate historical information even when $h(\cdot)$ is nonlinear in the spline coefficients.

When the mapping exhibits substantial curvature, the accuracy of this first-order Taylor approximation may deteriorate. In such cases, the normalizing constant $C(a)$ can be evaluated numerically using a Laplace approximation, adaptive quadrature, or Monte Carlo integration across a grid of $a$-values. These numerical strategies allow the NPP to incorporate complex external summaries--including nonlinear contrasts of spline-based treatment effects--while remaining computationally tractable. Although increasing spline richness or model dimensionality raises the computational burden of evaluating $C(a)$, approaches such as low-rank or penalized-spline bases, analytic marginalization of Gaussian components, or Laplace/INLA approximations can mitigate this burden while preserving the inferential gains of borrowing historical information on high-dimensional or functional treatment effects.

\subsection{Trial Design}
\label{s:trial_design}
In many enrichment or precision-medicine settings, a key aim is to identify the effective subspace, or the subgroup of patients for whom treatment is believed to provide clinically meaningful benefit. This corresponds to the set of covariate values where the treatment effect is most likely to exceed a prespecified threshold. For example, consider a study evaluating whether PAP therapy for OSA is more effective among patients with high versus low HB. In this context, the effective subspace would consist of the level(s) of HB for which there is high posterior probability that the treatment effect of PAP is sufficiently large to be clinically relevant. Let 
\begin{equation}
\gamma(x, t=1 \mid \boldsymbol{\beta})  
= \beta_2 + \beta_3 x
\end{equation}
denote the variable-specific treatment effect or blip function evaluated at $t=1$ \citep{robins1997causal}; for notational convenience we write $\gamma(x)$ throughout. The effective subspace is then defined as
\begin{equation}
\label{eq:eff_subspace}
    \mathcal{X}^* = \left\{ x : P\!\left( \gamma(x) > e_1 \mid \mathcal{D}\right) > 1-\alpha \right\};
\end{equation}
i.e., the set of covariate values $x$ for which there is strong posterior evidence that the treatment effect exceeds the clinically meaningful threshold $e_1$. In practice, $e_1$ is chosen based on subject-matter expertise, while the tuning parameter $\alpha$ is calibrated through simulation to achieve desirable operating characteristics. The effective subspace may consist of a continuous region or of multiple disjoint covariate-defined subgroups in which the treatment is predicted to be beneficial. A trial designed to detect heterogeneous treatment effects therefore targets the hypothesis
\begin{equation}
H_0: \gamma(x) \le e_1 \;\; \forall \;\; x \in \mathcal{X}  
\quad \text{vs.} \quad 
H_A: \exists \;\; x \in \mathcal{X} \;\; \text{s.t. } \gamma(x) > e_1,
\end{equation}
where $\mathcal{X}$ denotes the full space of candidate covariate values. Under the null hypothesis, the treatment fails to achieve the clinically relevant effect for every subgroup; under the alternative, there exists at least one subgroup--i.e., a nonempty effective subspace--for which the treatment meaningfully improves outcomes.

\subsection{Interim Analysis Procedure}
\label{s:interim}

At interim analysis $\ell$, let $\mathcal{D}_\ell$ denote the data accrued up to that look. 
The accumulating trial data are combined with external summary information via the NPP to update the posterior distribution of the treatment effect function $\gamma(x)$. The procedure is as follows:
\begin{enumerate}
        \item \textbf{Identify the effective subspace}: The model fitting procedure is based on the data observed up to the given interim analysis point, $\mathcal{D}_\ell$. Identify the effective subspace, $\mathcal{X}_\ell^*$, at analysis $\ell$:
        \begin{equation}
        \mathcal{X}^*_\ell = \{x :  P\!\left(\gamma(x) > e_1 \mid \mathcal{D}_\ell\right) > 1-\alpha\},
        \end{equation}
        If no biomarker level satisfies this criterion, treat the entire biomarker space as the effective subspace.  We then compute the enriched treatment effect at the $\ell$-th interim analysis: $\Delta_\ell = \int_{\mathcal{X}_\ell^*} \gamma(x) \widehat{f}_\ell(x)  \partial x,$
        where $\widehat{f}_\ell$ is the joint empirical probability distribution function of the predictive variables in the effective subspace as of that interim point in the trial.
        \item \textbf{Stop for efficacy}: Assess whether to stop the trial based on the posterior probability of trial success in the effective subspace using the following criterion: $P\left(\Delta_\ell>b_1\mid \mathcal{D}_\ell\right) > B_{1}.$
        \item \textbf{Stop for futility}: Assess whether to stop the trial based on the posterior probability of trial success in the effective subspace using the following criterion: $P\left(\Delta_\ell<b_2\mid \mathcal{D}_\ell\right) > B_{2}.$
        \item \textbf{Enrich the sample}: If no stopping boundary is crossed, accrual continues but only within $x \in \mathcal{X}^*_\ell$ until the next interim analysis. 
\end{enumerate}

The process repeats until either an early stopping criterion is met or the maximum planned sample size is reached. { The design requires the pre-specification of several quantities prior to trial initiation. The number and timing of interim analyses should balance the opportunity for early stopping against the precision of interim estimates; in practice, logistical constraints typically limit the number of interim analyses to 1-3 and a natural timing for the first interim is after approximately one-third to one-half of the maximum sample size has been enrolled \citep{berry2010bayesian}. Sample size determination for Bayesian adaptive designs is an active area of methodological research; in practice, a natural starting point is a frequentist sample size calculation for the analogous fixed design, refined through simulation by iterating over candidate maximum sample sizes and evaluating operating characteristics across a range of data-generating scenarios \citep{berry2010bayesian}. Practical guidance for selecting all hyperparameters, together with recommended examples, is provided in Web Appendix~D.}

\begin{table}[]
\centering
\footnotesize
\caption{Summary mappings for incorporating historical marginal treatment
effects into a current regression model. The historical study reports a
marginal treatment effect under link $g_{\text{hist}}$. The current trial
uses link $g_{\text{curr}}$ with conditional linear predictor
$\eta(t,x)=\beta_0+\beta_1 x+\beta_2 t+\beta_3 tx$ and
$\eta_t(x)\equiv\eta(t,x)$. The historical covariate distribution satisfies
$P(X=1)=\mu_X$. For rows 2--5, let $p_{tx}=\mathrm{logit}^{-1}(\eta_t(x))$
denote the fitted probability under $g_{\text{curr}}=\mathrm{logit}$, with
derivative $p_{tx}'=p_{tx}(1-p_{tx})$, marginal probability
$P_t=(1-\mu_X)\,p_{t0}+\mu_X\,p_{t1}$, and marginal derivative
$P_t'=(1-\mu_X)\,p_{t0}'\,v_{t0}+\mu_X\,p_{t1}'\,v_{t1}$, where
$v_{tx}=\partial\eta_t(x)/\partial\boldsymbol{\beta}^\top=(1,\,x,\,t,\,tx)$.
For each link pairing, $h(\boldsymbol{\beta})$ is the historical estimand
implied by the current model and $J(\boldsymbol{\beta})=\partial
h(\boldsymbol{\beta})/\partial\boldsymbol{\beta}^\top$ is its Jacobian.}
\label{tab:mapping-table}
\begin{tabular}{llll}
\toprule
\multicolumn{2}{c}{\textbf{Link}} && \\
$g_{\text{hist}}$ &
$g_{\text{curr}}$ &
\textbf{Mapping $h(\boldsymbol{\beta})$} &
\textbf{Jacobian $J(\boldsymbol{\beta})$} \\
\midrule
\addlinespace[6pt]

%-----------------------------------------
% 1. Identity -- Identity
%-----------------------------------------
Identity &
Identity &
$h(\boldsymbol{\beta})=\beta_2+\mu_X\beta_3$ &
$J(\boldsymbol{\beta})=
\begin{pmatrix} 0 & 0 & 1 & \mu_X \end{pmatrix}$
\\
\addlinespace[10pt]

%-----------------------------------------
% 2. Identity -- Logit
%-----------------------------------------
Identity &
Logit &
$h(\boldsymbol{\beta})=
(1-\mu_X)(p_{10}-p_{00})+\mu_X(p_{11}-p_{01})$ &
$J(\boldsymbol{\beta})=
(1-\mu_X)(p_{10}'\,v_{10}-p_{00}'\,v_{00})
+\mu_X(p_{11}'\,v_{11}-p_{01}'\,v_{01})$
\\
\addlinespace[10pt]

%-----------------------------------------
% 3. Logit -- Logit
%-----------------------------------------
Logit &
Logit &
$h(\boldsymbol{\beta})=
\mathrm{logit}(P_1)-\mathrm{logit}(P_0)$ &
$J(\boldsymbol{\beta})=
\dfrac{P_1'}{P_1(1-P_1)}
-\dfrac{P_0'}{P_0(1-P_0)}$
\\
\addlinespace[10pt]

%-----------------------------------------
% 4. Log -- Logit
%-----------------------------------------
Log &
Logit &
$h(\boldsymbol{\beta})=
\log(P_1)-\log(P_0)$ &
$J(\boldsymbol{\beta})=
\dfrac{P_1'}{P_1}
-\dfrac{P_0'}{P_0}$
\\
\addlinespace[10pt]

%-----------------------------------------
% 5. Inverse -- Logit
%-----------------------------------------
Inverse &
Logit &
$h(\boldsymbol{\beta})=
P_1^{-1}-P_0^{-1}$ &
$J(\boldsymbol{\beta})=
-\dfrac{P_1'}{P_1^2}
+\dfrac{P_0'}{P_0^2}
$
\\
\addlinespace[6pt]

\bottomrule
\end{tabular}
\end{table}

\section{Simulation Study}
\label{sec:sims}

We evaluated the proposed adaptive enrichment design with NPP-based borrowing  in a binary-outcome setting where historical summaries are reported on a marginal 
log-odds ratio scale (i.e., the logit-logit mapping in Table~\ref{tab:mapping-table}), 
inducing a nonlinear and non-collapsible mapping to the conditional regression 
parameters. { As no existing method accommodates summary-level historical borrowing, we benchmark against a no-borrowing design ($n_t = 0$), which represents the operating characteristics a practitioner would achieve by ignoring the available historical information entirely.} Specifically, outcomes were generated from the logistic model
\begin{equation}
\text{logit}\{P(Y_i = 1 \mid X_i, \mathrm{T}_i)\}
= \beta_0 + \beta_1 X_i + \beta_2\,\mathrm{T}_i + \beta_3\,\mathrm{T}_i X_i,
\end{equation}
with $(\beta_0,\beta_1,\beta_2)=(-0.2,\,0.4,\,0)$ and independent treatment and biomarker indicators with $P(X_i=1)=P(\mathrm{T}_i=1)=0.5$. The conditional treatment effect (blip function) was $\gamma(x) = \beta_2 + \beta_3 x$, and treatment effectiveness at covariate value $x$ was defined relative to the threshold $e_1 = 0$. We considered two configurations for effect modification: a null setting where treatment is ineffective for all patients ($\beta_3=0$) and an alternative ($\beta_3=0.65$), where treatment is effective in only individuals with $X = 1$.  Each simulated trial enrolled up to $n=600$ participants, with one interim analysis at $n=400$. Posterior updating, effective subspace construction, and decision-making followed Section~\ref{s:interim}, with early stopping for efficacy if $P(\Delta_\ell>0\mid \mathcal D_\ell)>0.99$ and for futility if $P(\Delta_\ell<0\mid \mathcal D_\ell)>0.80$; if neither criterion was met, accrual continued within the currently identified effective subspace $\mathcal X_\ell^*$ ($\alpha=0.05$ and $e_1=0$). 

External information was incorporated through the NPP from a single historical marginal log-odds ratio. Let $\mu_X^{\mathrm{hist}}$ denote the historical biomarker prevalence and define
\begin{equation}
p_{tx}=\mathrm{logit}^{-1}(\beta_0+\beta_1 x+\beta_2 t+\beta_3 tx), \qquad t,x\in\{0,1\},
\end{equation}
with marginal risks $P_1=(1-\mu_X^{\mathrm{hist}})p_{10}+\mu_X^{\mathrm{hist}}p_{11}$ and $P_0=(1-\mu_X^{\mathrm{hist}})p_{00}+\mu_X^{\mathrm{hist}}p_{01}$. The implied historical summary is the marginal log-odds ratio
\begin{equation}
h(\boldsymbol{\beta})=\text{logit}(P_1)-\text{logit}(P_0).
\end{equation}
{
We encode historical information through the estimated average treatment effect $\hat{\Delta}$ from the historical study, where $\hat{\Delta} = h(\boldsymbol{\beta}) 
+ \delta$ is the observed summary statistic, $h(\boldsymbol{\beta})$ is the true historical estimand implied by the current model, $\delta$ controls prior-data conflict, and $s_\Delta^2$ is the variance of the estimator $\hat{\Delta}$, derived via the delta method as}
\begin{equation}
s_\Delta^2=\frac{1}{n_t\,P_1(1-P_1)}+\frac{1}{n_c\,P_0(1-P_0)},
\end{equation}
thereby reverse-engineering a hypothetical historical study with sample sizes $(n_t,n_c)$ and covariate distribution $\mu_X^{\mathrm{hist}}$. The baseline prior for all coefficients was $\boldsymbol{\beta}\sim\mathcal N(\mathbf 0,5^2\mathbf I)$ and the borrowing weight followed $a\sim \mathrm{Beta}(4,1)$.

Because $h(\boldsymbol{\beta})$ is nonlinear, the NPP normalizing constant $C(a)$ is unavailable in closed form. We compared two implementations: (i) the proposed \emph{Linearized} approach, which uses a first-order Taylor expansion of $h(\boldsymbol{\beta})$ as in Section~\ref{s:nonlinear} and reuses the closed-form expression for $C(a)$ from the linear case; and (ii) a \emph{Exact} approach that evaluates the exact mapping within the sampler and approximates $C(a)$ by Monte Carlo integration under the baseline prior \citep{ye_normalized_2022}. Scenarios varied the historical sample size $n_t=n_c\in\{300,500,700\}$, historical bias $\delta\in\{-0.1,0,0.1\}$, and historical biomarker prevalence $\mu_X^{\mathrm{hist}}\in\{0.3,0.5,0.7\}$; each configuration was replicated 1{,}000 times. Posterior inference used \texttt{Stan} via \texttt{rstan}, with one chain of 2{,}500 iterations (500 warm-up). For the Exact method, $\log C(a)$ was precomputed on a grid of $a$ values using $M=20{,}000$ prior draws and interpolated during sampling. An additional simulation study with a wider range of \(\delta\) is described in Web Appendix~E.

We summarize operating characteristics across replicates as follows. The traditional Type~I error is the empirical probability of incorrectly stopping for efficacy under $\beta_3 = 0$, while power is the probability of correctly stopping for efficacy under $\beta_3 = 0.65$. Following \cite{chapple2018subgroup}, generalized power is defined as the empirical probability that (1) the efficacy criterion is met at either an interim or final analysis and (2) the correct effective subspace is identified. We also report the posterior mean borrowing weight $\mathbb{E}[a \mid \mathcal{D}]$, mean sample size at trial completion (expected sample size [ESS]), and the proportion of early stopping for futility.

Tables~\ref{tab:typeI} and~\ref{tab:power} summarize the simulation results under the null and alternative configurations respectively. Positive values of $\delta$ correspond to optimistic historical evidence (larger marginal odds ratios), while negative values correspond to pessimistic or harmful historical signals. Across all scenarios, the mean posterior borrowing weight $\mathbb{E}[a\mid\mathcal{D}]$ remained close to $0.75$--$0.80$, indicating that the range of bias values considered was not large enough to trigger meaningful contraction of the borrowing weight. The empirical Type~I error remained well controlled below the nominal $0.05$ level, and borrowing substantially improved performance relative to the no-borrowing benchmark: generalized power increased from $0.69$ to $0.76$--$0.94$ and expected sample size reduced by an average of $43$ patients. Across all scenarios, the Linearized and Exact implementations produced nearly indistinguishable operating characteristics, indicating the finite-sample accuracy of the Taylor approximation. Web Appendix E reports operating characteristics across a wider range of historical bias values.

\begin{table}[htbp]
\centering
\caption{Operating characteristics under $\beta_3 = 0$ (Type I error).
The column ``Borrow Method" distinguishes our Linearized approach, which uses a first-order Taylor expansion and a closed-form NPP normalizing constant, from the Exact implementation, which repeatedly computes the exact logit-logit mapping within the sampler and a Monte Carlo-based normalizing constant. Shaded rows are duplicated intentionally to allow direct comparison across changes in $\delta$ (holding $n_t$ and $\mu_X^{hist}$ fixed) and across changes in $\mu_X^{hist}$ (holding $n_t$ and $\delta$ fixed).}
\label{tab:typeI}
\begin{tabular}{llllcccc}
\toprule
$n_t$ & $\delta$ & $\mu_X^{hist}$ & \textbf{Borrow Method} & $\mathbb{E}[a \mid \mathcal{D}]$ & \textbf{Type I Error} & \textbf{Futility} & \textbf{ESS} \\
\midrule
\multicolumn{8}{l}{\textit{No historical data}} \\
0   &  - & - & - & - & 0.033 & 0.27 & 553.1 \\
\multicolumn{8}{l}{\textit{With historical borrowing}} \\
\multicolumn{8}{l}{\footnotesize Vary $n_t$ at $\delta = 0$, $\mu_X^{hist} = 0.5$} \\
% Block A: vary n_t at delta = 0, mu_X = 0.5
300 & 0.0 & 0.5 & Linearized & 0.81 & 0.016 & 0.16 & 578.2 \\ 
  300 & 0.0 & 0.5 & Exact & 0.80 & 0.017 & 0.16 & 577.2 \\ 
  500 & 0.0 & 0.5 & Linearized & 0.80 & 0.016 & 0.11 & 584.8 \\ 
  500 & 0.0 & 0.5 & Exact & 0.80 & 0.020 & 0.12 & 583.8 \\ 
  700 & 0.0 & 0.5 & Linearized & 0.80 & 0.017 & 0.09 & 588.4 \\ 
  700 & 0.0 & 0.5 & Exact & 0.80 & 0.021 & 0.09 & 587.6  \\[2pt]
\multicolumn{8}{l}{\footnotesize Vary $\delta$ at $n_t = 500$, $\mu_X^{hist} = 0.5$} \\
% Block B: vary delta at n_t = 500, mu_X = 0.5
   500 & -0.1 & 0.5 & Linearized & 0.80 & 0.009 & 0.43 & 528.6 \\ 
  500 & -0.1 & 0.5 & Exact & 0.80 & 0.006 & 0.44 & 529.0 \\ 
  \rowcolor{gray!15}500 & 0.0 & 0.5 & Linearized & 0.80 & 0.016 & 0.11 & 584.8 \\ 
  \rowcolor{gray!15}500 & 0.0 & 0.5 & Exact & 0.80 & 0.020 & 0.12 & 583.8 \\ 
   500 & 0.1 & 0.5 & Linearized & 0.80 & 0.042 & 0.02 & 593.6 \\ 
  500 & 0.1 & 0.5 & Exact & 0.80 & 0.043 & 0.02 & 593.2 \\ 
% Block C: vary mu_X at n_t = 500, delta = 0
\multicolumn{8}{l}{\footnotesize Vary $\mu_X^{hist}$ at $n_t = 500$, $\delta = 0$} \\
500 & 0.0 & 0.3 & Linearized & 0.80 & 0.018 & 0.17 & 570.8 \\ 
  500 & 0.0 & 0.3 & Exact & 0.80 & 0.016 & 0.18 & 570.8 \\ 
 \rowcolor{gray!15}500 & 0.0 & 0.5 & Linearized & 0.80 & 0.016 & 0.11 & 584.8 \\ 
  \rowcolor{gray!15}500 & 0.0 & 0.5 & Exact & 0.80 & 0.020 & 0.12 & 583.8 \\ 
  500 & 0.0 & 0.7 & Linearized & 0.80 & 0.012 & 0.18 & 573.6 \\ 
  500 & 0.0 & 0.7 & Exact & 0.80 & 0.019 & 0.17 & 573.2 \\ 
\bottomrule
\end{tabular}
\begin{flushleft}
\footnotesize
\textit{Notes.} 
The first row (\textit{No historical data}) represents the comparator design without borrowing ($n_t=0$). $\mathbb{E}[a|\mathcal{D}]$ is the mean posterior borrowing weight, and ESS denotes the expected sample size at trial completion.
\end{flushleft}
\end{table}

%Although runtimes were similar in this low-dimensional logistic regression setting, the Linearized formulation remains important for both methodological and practical reasons. By yielding a closed-form normalizing constant for the NPP, this approach avoids Monte Carlo precomputation and interpolation, which scale poorly as model complexity increases. Moreover, the analytic form of the linearized mapping makes the relationship between historical summaries and current model parameters explicit, facilitating sensitivity analyses with respect to historical effect size, uncertainty, and covariate prevalence--tasks that would otherwise require repeated refitting of a fully nonlinear model. 

\begin{table}[htbp]
\centering
\caption{Operating characteristics under $\beta_3 = 0.65$ (Power).
The column ``Borrow Method" distinguishes our Linearized approach, which uses a first-order Taylor expansion and a closed-form NPP normalizing constant, from the Exact implementation, which repeatedly computes the exact logit-logit mapping within the sampler and uses a Monte Carlo-based normalizing constant. Shaded rows are duplicated intentionally to allow direct comparison across changes in $\delta$ (holding $n_t$ and $\mu_X^{hist}$ fixed) and across changes in $\mu_X^{hist}$ (holding $n_t$ and $\delta$ fixed).}
\label{tab:power}
\begin{tabular}{llllccccc}
\toprule
$n_t$ & $\delta$ & $\mu_X^{hist}$ & \textbf{Borrow Method} & $\mathbb{E}[a \mid \mathcal{D}]$ & \textbf{Power} & \textbf{Gen. Power} & \textbf{Futility} & \textbf{ESS} \\
\midrule
\multicolumn{8}{l}{\textit{No historical data}} \\
0   &  - & - & - & - & 0.73 & 0.69 & 0.01 & 503.4 \\
\multicolumn{8}{l}{\textit{With historical borrowing}} \\
\multicolumn{8}{l}{\footnotesize Vary $n_t$ at $\delta = 0$, $\mu_X^{hist} = 0.5$} \\
% Block A: vary n_t at delta = 0, mu_X = 0.5
300 & 0.0 & 0.5 & Linearized & 0.81 & 0.86 & 0.82 & 0.00 & 473.0 \\ 
  300 & 0.0 & 0.5 & Exact & 0.81 & 0.86 & 0.82 & 0.00 & 472.6 \\ 
  500 & 0.0 & 0.5 & Linearized & 0.80 & 0.90 & 0.85 & 0.00 & 463.6 \\ 
  500 & 0.0 & 0.5 & Exact & 0.81 & 0.91 & 0.85 & 0.00 & 461.2 \\ 
  700 & 0.0 & 0.5 & Linearized & 0.80 & 0.93 & 0.86 & 0.00 & 452.2 \\ 
  700 & 0.0 & 0.5 & Exact & 0.80 & 0.93 & 0.86 & 0.00 & 450.8 \\ [2pt]

% Block B: vary delta at n_t = 500, mu_X = 0.5
\multicolumn{8}{l}{\footnotesize Vary $\delta$ at $n_t = 500$, $\mu_X^{hist} = 0.5$} \\
   500 & -0.1 & 0.5 & Linearized & 0.80 & 0.82 & 0.78 & 0.00 & 486.8 \\ 
  500 & -0.1 & 0.5 & Exact & 0.80 & 0.83 & 0.79 & 0.00 & 484.2 \\ 
  \rowcolor{gray!15}500 & 0.0 & 0.5 & Linearized & 0.80 & 0.90 & 0.85 & 0.00 & 463.6 \\ 
  \rowcolor{gray!15}500 & 0.0 & 0.5 & Exact & 0.81 & 0.91 & 0.85 & 0.00 & 461.2 \\ 
   500 & 0.1 & 0.5 & Linearized & 0.80 & 0.94 & 0.86 & 0.00 & 439.0 \\ 
  500 & 0.1 & 0.5 & Exact & 0.80 & 0.94 & 0.86 & 0.00 & 438.4 \\ 

% Block C: vary mu_X at n_t = 500, delta = 0
\multicolumn{8}{l}{\footnotesize Vary $\mu_X^{hist}$ at $n_t = 500$, $\delta = 0$} \\
500 & 0.0 & 0.3 & Linearized & 0.80 & 0.79 & 0.76 & 0.00 & 494.0 \\ 
  500 & 0.0 & 0.3 & Exact & 0.80 & 0.79 & 0.77 & 0.00 & 495.0 \\ 
  \rowcolor{gray!15}500 & 0.0 & 0.5 & Linearized & 0.80 & 0.90 & 0.85 & 0.00 & 463.6 \\ 
  \rowcolor{gray!15}500 & 0.0 & 0.5 & Exact & 0.81 & 0.91 & 0.85 & 0.00 & 461.2 \\ 
  500 & 0.0 & 0.7 & Linearized & 0.80 & 0.98 & 0.93 & 0.00 & 415.8 \\ 
  500 & 0.0 & 0.7 & Exact & 0.80 & 0.99 & 0.94 & 0.00 & 412.4 \\ 
\bottomrule
\end{tabular}
\begin{flushleft}
\footnotesize
\textit{Notes.} 
The first row (\textit{No historical data}) represents the comparator design without borrowing ($n_t=0$). $\mathbb{E}[a|\mathcal{D}]$ is the mean posterior borrowing weight, gen. power is generalized power, and ESS denotes the expected sample size at trial completion.
\end{flushleft}
\end{table}

Web Appendix F reports two additional simulation studies under the identity link. The first evaluates operating characteristics across historical bias, biomarker prevalence, and historical sample size, while the second examines sensitivity to prior informativeness through $a \sim \mathrm{Beta}(\eta,1)$ with $\eta \in \{1,4\}$ over a dense grid of historical bias values. Under the null configuration ($\beta_3=0$), the no-borrowing design achieved nominal Type~I error, and NPP-based borrowing generally preserved error control when historical information was unbiased or pessimistic, with empirical Type~I error ranging from 0.012 to 0.072 across borrowing scenarios; inflation was observed primarily when the historical ATE was positively biased and borrowing was strongly favored. Under the heterogeneous-effect configuration, borrowing increased power from 0.68 to as high as 0.94 and generalized power from 0.62 to 0.88, with gains largest when historical data were concordant and biomarker prevalence was high. Across both studies, increased historical compatibility led to earlier stopping, reduced ESS, and improved correct identification of the effective subspace, while sensitivity analyses confirmed the expected trade-off between prior informativeness and robustness to historical bias.

\section{Obstructive Sleep Apnea Trial with Historical Borrowing}
\label{sec:data_example}
To illustrate the proposed adaptive enrichment design, we consider a hypothetical randomized controlled trial in OSA that incorporates external information on the ATE through a NPP. The objective of the trial is to evaluate whether patients at higher cardiovascular risk, as identified by elevated HB, experience greater improvements in intermediate cardiovascular outcomes following PAP therapy compared with lower-risk patients during a 6-month intervention period. The primary outcome is change in 24-hour systolic blood pressure.

Hypoxic burden was selected based on previous work ~\citep{azarbarzin2019hypoxic,azarbarzin2021sleep}, which identified oximetry-derived measures of hypoxemia as strong predictors of cardiovascular morbidity, mortality, and response to PAP therapy. Unlike the conventional apnea-hypopnea index, HB captures the cumulative physiological burden of hypoxemia. Despite these advances, several large randomized trials (SAVE, RICCADSA, and ISAAC) reported neutral effects of PAP therapy on major adverse cardiovascular events \citep{sanchez2020effect,peker2016effect,mcevoy2016cpap}, in part because they did not account for heterogeneity in biomarker-defined risk. Only SAVE and ISAAC reported treatment effects on systolic blood pressure. In SAVE (ClinicalTrials.gov: NCT00738179), adults aged 45-75 years with moderate-to-severe OSA and established cardiovascular disease were randomized to CPAP plus usual care or usual care alone, yielding an estimated between-arm difference of $-0.4$ mmHg (95\% CI: $-1.5,\,0.8$) after a mean follow-up of 3.7 years ($n_t=1166$, $n_c=1158$). In ISAAC (ClinicalTrials.gov: NCT01335087), conducted among patients with acute coronary syndrome at 15 Spanish hospitals, the corresponding estimate was $0.07$ mmHg (95\% CI: $-2.94,\,3.09$) after 48 months ($n_t=554$, $n_c=539$).

We evaluate the operating characteristics of the proposed design through simulation under two biomarker strata, low HB $(X=0)$ and high HB $(X=1)$, where a favorable treatment effect corresponds to a reduction in systolic blood pressure. Outcomes are generated from the working model $
    Y_i = \beta_0 + \beta_1 X_i + \beta_2 \,\mathrm{T}_i + \beta_3 \,\mathrm{T}_i X_i + \varepsilon_i$, $\varepsilon_i \sim \mathcal{N}(0,1)$, so that the treatment effect at biomarker level $x$ is $\gamma(x)=\beta_2+\beta_3 x$. We assume equal randomization to PAP and control arms, a 50\% prevalence of high HB, and true parameter values $\beta_0=\beta_1=0$. Two data-generating configurations are considered: a null scenario where treatment is ineffective for the entire sample with $(\beta_2,\beta_3)=(0,0)$, and an alternative scenario with $(\beta_2,\beta_3)=(0.47,-0.94)$, under which the high-HB subgroup experiences a clinically meaningful benefit. Based on prior OSA trials, the pooled standard deviation of systolic blood pressure is approximately 8.5 mmHg; for interpretability, outcomes and regression coefficients are rescaled so that the marginal standard deviation of $Y_i$ is 1. Under this scaling, $(\beta_2,\beta_3)=(0.47,-0.94)$ corresponds on the original scale to a clinically meaningful treatment effect of approximately -4 mmHg in the high-HB subgroup. The outcome is centered so that the intercept is zero.

Historical information from SAVE and ISAAC is incorporated through their reported ATEs of $\hat{\Delta}=(-0.40,\,0.07)$ and associated variances $V_h=(0.597,\,1.538)$, assuming a historical biomarker prevalence of $\mu_X^{\mathrm{hist}}=0.5$. Under the current model, the implied historical ATE is $\theta=\beta_2+\mu_X^{\mathrm{hist}}\beta_3$. Borrowing follows the NPP framework described in Section~\ref{s:npp}, where each historical dataset contributes a discounted likelihood term $\mathcal{N}(\hat{\delta}_h \mid \theta, V_h)^{a_h}$ with study-specific borrowing weights $a_h \sim \mathrm{Beta}(4,1)$. Independent $\mathcal{N}(0,5^2)$ priors are assigned to $(\beta_1,\beta_2,\beta_3)$, and the residual variance follows $\sigma^2 \sim \mathrm{IG}(2,2)$. Posterior inference is conducted under two scenarios: a borrowing analysis with $(a_1,a_2)$ estimated from the data, and a no-borrowing analysis with $a_1=a_2=0$.

The trial enrolls up to 300 participants with equal randomization and a single interim analysis at $n=200$, using the interim procedure described in Section~\ref{s:interim}. Because a favorable treatment effect corresponds to a decrease in systolic blood pressure, all stopping and subgroup-identification criteria are applied in the negative direction of the estimated treatment effects, which is implemented by reversing the sign of the estimated effects prior to applying the decision rules. All thresholds and MCMC settings match those used in the main simulation study (Section~\ref{sec:sims}), except that $B_1=0.975$.

Table~\ref{tab:osa_enrichment} summarizes the operating characteristics of the adaptive enrichment design under both data-generating scenarios. Under the null scenario, borrowing reduces the Type I error from 0.06 to 0.01 and increases the ESS from 277.5 to 285.2. This occurs because, in the borrowing setting, estimated treatment effects--whether at the subgroup or overall level--are shrunk toward the historical ATEs, which favor a null average effect. In contrast, the no-borrowing approach is more reactive to random extremes in the observed data, leading to increased Type I error when effects are spuriously high and a higher probability of stopping for futility when effects are spuriously low. Under the heterogeneous-effect scenario, borrowing increases both power and generalized power from 0.77 to 0.90 and reduces the futility rate, indicating earlier and more reliable identification of the truly effective subgroup.

\begin{table}[htbp]
\centering
\caption{Operating characteristics of the OSA adaptive enrichment design under null (\(\beta_3 = 0\)) and alternative (\(\beta_3 = -0.94\)) interaction scenarios, with and without historical borrowing. The “Power'' column corresponds to Type I error (T1E) when \(\beta_3 = 0\) and to frequentist power when \(\beta_3 = -0.94\).}
\label{tab:osa_enrichment}
\begin{tabular}{lcccccc}
\toprule

($\beta_2,\beta_3$)  & ${\mathbb{E}[a_1 \mid \mathcal{D}]}$ & ${\mathbb{E}[a_2 \mid \mathcal{D}]}$& \textbf{Power} & \textbf{Generalized Power} & \textbf{Futility} & \textbf{ESS} \\
\midrule
\multicolumn{7}{l}{\textit{No historical data}} \\
$(0,0)$ & - & - & 0.06 (T1E) & - & 0.25 & 277.5 \\ 
 $(0.47,-0.94)$ & - & - & 0.77 & 0.77 & 0.16 & 229.7 \\ 
\multicolumn{7}{l}{\textit{With historical borrowing}} \\
  $(0,0)$ & 0.80 & 0.79 & 0.01 (T1E) & - & 0.21 & 285.2 \\ 
  $(0.47,-0.94)$ & 0.80 & 0.80 & 0.90 & 0.90 & 0.06 & 222.0 \\ 
\bottomrule
\end{tabular}
\begin{flushleft}
\footnotesize
\textit{Notes.} The first two rows (\textit{No historical data}) represent the comparator model without borrowing ($n_t=0$).  
$\mathbb{E}[a_h\mid \mathcal{D}]$ is the mean posterior borrowing weight for study $h$, and ESS denotes the expected sample size at trial completion.
\end{flushleft}
\end{table}

To provide a concrete illustration of the interim decision-making process, we consider a single simulated trial under the heterogeneous-effect scenario at the first interim analysis (\(n = 200\)). The data were generated under the same model and parameter configuration as above. We fit both the borrowing and no-borrowing models to the interim data in \texttt{rstan} using four chains, and 2{,}500 iterations per chain with  1,000 warm-up iterations; MCMC diagnostics indicated satisfactory convergence (Web Appendix G). Here, both models recovered the expected pattern of treatment effects of minimal benefit in low-HB patients and meaningful reductions in systolic blood pressure for high-HB patients. Under borrowing, the posterior mean treatment effect was approximately 2.7 mmHg (95\% CrI: 0.35 to 5.1) in the low-HB subgroup and -2.6 mmHg (95\% CrI: -5.1 to -0.15) in the high-HB subgroup, with credible intervals 25-30\% narrower than in the no-borrowing model. In contrast, the no-borrowing fit yielded posterior means of about 3.4 mmHg (95\% CrI: 0.44 to 6.3) in the low-HB subgroup and -1.9 mmHg (95\% CrI: -5.1 to 1.2) in the high-HB subgroup, with the latter interval still crossing zero.
\section{Discussion}
\label{sec:discussion}
This work introduced an adaptive enrichment design that borrows information from external studies through a NPP anchored on summary measures derived from historical data. A major advantage of this approach is that it can incorporate aggregate-level summaries that are functions of model parameters, rather than requiring access to individual participant data or to parameter-specific external estimates. { The proposed summary-anchored NPP is the only existing framework, to our knowledge, that enables principled adaptive borrowing from published aggregate summaries in a subgroup-finding enrichment trial.  The simulations demonstrated that the proposed approach can substantially improve power and generalized power relative to a no-borrowing benchmark, particularly when historical and current populations are well aligned, while maintaining acceptable Type~I error control under mild bias.}

{

A further consideration arises when historical information is reported on a non-collapsible scale, such as the marginal odds ratio common in RCTs. Our summary-anchored NPP framework operates directly in this setting by introducing a linearized logit-logit mapping, obtained via a first-order Taylor expansion, which yields a closed-form normalizing constant and a transparent borrowing mechanism. We compare this approach with a nonlinear alternative that recomputes the exact mapping within the sampler using a Monte Carlo normalizing constant. Although runtimes were similar in our four-parameter example, the linearized method avoids repeated Monte Carlo integration in sensitivity analyses, provides an explicit link between historical and current parameters, and scales more naturally to higher-dimensional or hierarchical models where iterative integration may be unstable. In simulations, the two approaches produced nearly identical operating characteristics--Type I error was well controlled in all scenarios, and generalized power differed by at most 0.009--while both substantially outperformed the no-borrowing benchmark.
}

The proposed framework belongs to the broader family of dynamic borrowing priors, including commensurate priors, robust MAP priors, and hierarchical mixture models. These methods can accommodate external information when it is available directly on the parameters of the current model---for example, as regression coefficients or log-hazard ratios. However, when the available evidence consists of under-identified summary measures, such as a marginal ATE when the current model targets subgroup-specific effects, these methods cannot be applied without non-trivial reparameterization. The proposed summary-anchored NPP overcomes this limitation by enabling borrowing through any user-specified function of the parameters while relying only on published aggregate information, making it particularly well suited to modern trial settings where individual-level data are inaccessible.

{

The OSA case study illustrates how the proposed framework can be extended to incorporate evidence from more than one historical trial. In this example, we summarized each study by its published between-arm difference in systolic blood pressure and its reported uncertainty, and then allowed the design to learn a separate borrowing weight for each study. These study-specific weights were estimated jointly with the current-trial parameters, enabling the method to borrow more heavily from studies that align with the current data and less from those that did not. Operationally, this provides a flexible and transparent extension of the summary-anchored approach to multiple external sources-an important advantage in therapeutic areas, such as OSA, where several moderate-sized trials exist but individual-level data remain inaccessible. In the data example, borrowing from two historical studies increased generalized power from 0.77 to 0.90, while maintaining Type I error. 
}

Several limitations warrant discussion. First, {as in all existing borrowing methods}, the borrowing weight does not sufficiently contract under subtle positive bias in the external data, leading to modest Type I error inflation; { addressing this through truncation mechanisms or LASSO-type shrinkage on the borrowing weight is an interesting direction for future research.} Second, the design assumes that the historical summary variance accurately reflects uncertainty in the external estimate. Underestimated variances or unrecognized heterogeneity across studies could lead to over-borrowing and inflated posterior confidence. Third, the posterior mean borrowing weight remained moderate even when the historical and current data were highly concordant, reflecting the conservative nature of NPPs. This behavior aligns with known properties of NPPs, which tend to bias the borrowing weight downward \citep{shen2023optimal, pawel2023normalized}. Fixed or empirically estimated borrowing weights \citep{ibrahim2015power, gravestock2017adaptive} may offer practical alternatives, though at the cost of reduced robustness.

Future research will focus on extending the summary-anchored NPP to higher-dimensional biomarkers and longitudinal outcomes, as well as investigating formal decision-theoretic criteria for choosing hyperparameters and interim boundaries. Additional work is also needed to define regulatory-relevant criteria for Bayesian error control, distinguishing between Bayesian and frequentist notions of Type I error and power in adaptive contexts. Overall, this study provides a practical pathway for integrating historical evidence into Bayesian adaptive enrichment designs-strengthening inference in precision-medicine trials while maintaining transparency, interpretability, and rigorous evaluation of operating characteristics.

\section*{Supplementary Materials}

Web Appendices A-G, referenced in Sections~\ref{s:nonlinear}, 
\ref{s:interim}, \ref{sec:sims}, and \ref{sec:data_example}, will be available with the final version of this paper.

\section*{Acknowledgments}
Lara Maleyeff was supported by a Canadian Network for Statistical Training in Trials (CANSTAT) trainee award funded by Canadian Institutes of Health Research (CIHR) grant \#262556 for a portion of this research. Shirin Golchi is a Fonds de Recherche du Qu\'ebec, Sant\'e, Chercheuse-boursiere (Junior 1) and acknowledges support from a Natural Sciences and Engineering Research Council of Canada (NSERC) Discovery Grant, Canadian Statistical Sciences Institute (CANSSI), and the Fonds de Recherche du Qu\'ebec, Nature et technologies (FRQNT-NSERC NOVA). Erica E. M. Moodie is a Canada Research Chair (Tier 1) in Statistical Methods for Precision Medicine.

\newpage

\bibliographystyle{unsrt}  
\bibliography{main}  %%% Remove comment to use the external .bib file (using bibtex).

\end{document}